\documentclass[pra,amsmath,amssymb,onecolumn, 10pt]{revtex4}

\usepackage{amsthm}
\usepackage{dcolumn}
\usepackage{bm}
\usepackage{graphicx}

\begin{document}

\newtheorem{corollary}{Corollary}
\newtheorem{definition}{Definition}
\newtheorem{example}{Example}
\newtheorem{lemma}{Lemma}
\newtheorem{proposition}{Proposition}
\newtheorem{theorem}{Theorem}
\newtheorem{fact}{Fact}
\newtheorem{property}{Property}
\newcommand{\bra}[1]{\langle #1|}
\newcommand{\ket}[1]{|#1\rangle}
\newcommand{\braket}[3]{\langle #1|#2|#3\rangle}
\newcommand{\ip}[2]{\langle #1|#2\rangle}
\newcommand{\op}[2]{|#1\rangle \langle #2|}

\newcommand{\tr}{{\rm tr}}
\newcommand {\E } {{\mathcal{E}}}
\newcommand {\F } {{\mathcal{F}}}
\newcommand {\diag } {{\rm diag}}
\newcommand{\slocc}{\overset{\underset{\mathrm{SLOCC}}{}}{\longrightarrow}}

\title{\Large {\bf Multi-partite $W-$type state is determined by its single particle reduced density matrices}}
\author{Nengkun Yu}
\email{nengkunyu@gmail.com}
\affiliation{State Key Laboratory of Intelligent Technology and Systems, Tsinghua National Laboratory\protect\\
for Information Science and Technology, Department of Computer Science and Technology,\protect\\
Tsinghua University, Beijing 100084, China \\
Center for Quantum Computation
and Intelligent Systems (QCIS), Faculty of Engineering and
Information Technology, University of Technology, Sydney, NSW 2007,
Australia
}
\begin{abstract}
In this short note, we show that multi-partite $W$-type state is up to local unitaries uniquely determined by its reduced density matrices.
\end{abstract}

\maketitle

With entanglement being a proven asset to information
processing and computational tasks, much effort has been spent on quantifying it as a resource, and lots of results
have been obtained. It is extremely difficult to give a perfect description of the entanglement in multipartite case,
even it is well understood for bipartite pure states.

One important approach is to consider their interconvertibility through manipulations that do not require quantum
communication, which is to determine weather one can interconvert between two given states $\ket{\Psi}$ and $\ket{\Phi}$ by local
operation. A widely studied equivalence relation of multipartite state space is stochastic local operations and classical communication (SLOCC) \cite{EB01,HEB04,VDMV04}: $\ket{\Psi}$ and $\ket{\Phi}$ are considered
to be SLOCC equivalent if they can be reversibly converted from one to the other by operations belonging
to the class of stochastic local operations and classical communication. On the other hand, if only
local unitaries are allowed, the problem become very interesting and significant: local unitaries don't change entanglement, LU
equivalence states have the same entanglement(both for type and amount). Thus, LU equivalence relationship can be
considered as one key solution of characterization of multipartite entanglement. Two multipartite states $\ket{\Psi}$ and $\ket{\Phi}$
are called LU equivalence if there exist unitaries $U_1, \cdots,U_n$, such that $\ket{\Psi} = (U_1
\otimes
U_n)\ket{\Phi}$.
Generally, it is difficult to determine that whether two given multipartite states are LU equivalent, while it might be easier in the SLOCC case.

In this short note, we study a special case of the following interesting problem: how to check if two quantum states are LU equivalent, provided they lie in the same SLOCC class. For two $n-$partite pure states which are SLOCC equivalent to $\ket{W}_n=\frac{1}{\sqrt{n}}(\ket{0\cdots0 1}+\cdots+\ket{10\cdots 0})$, we show that they are LU equivalent if and only if they share the spectra of single particle reduced density matrices. This can be regarded as a stronger result of \cite{PR09} in which it was showed that n-qubit $W-$type state is determined by their bipartite reduced density matrices.

The following lemma gives a very nice characterization about $W$-type state up to local unitaries \cite{RP11,CCL11}.
\begin{lemma}
Any $W$-type pure state is LU equivalent to $$\sqrt{x}\ket{0\cdots 0}+\sqrt{c_1}\ket{0\cdots0 1}+\cdots+\sqrt{c_n}\ket{10\cdots 0},$$
with some $c_k>0$ and $x\geq 0$.
\end{lemma}
\textit{Proof}: Suppose $\ket{\psi}=(A_1\otimes A_2\otimes\cdots\otimes A_n)\ket{W}_n$ with $A_k$ being all non-singular 2-by-2 matrix.
For any $A_k$, there is a unitary $V_k$ such that $A_k=V_kB_k$ with $B_k$ being an upper triangle matrix. Thus, $\ket{\psi}$ is LU equivalent to
$$(B_1\otimes B_2\otimes\cdots\otimes B_n)\ket{W}_n=d\ket{0\cdots 0}+d_1\ket{0\cdots0 1}+\cdots+d_n\ket{10\cdots 0},$$
for complex $d,d_k$.
One can find diagonal unitary to transform it into the wanted formalism.
\hfill $\blacksquare$

Now we can present our main result as follows
\begin{theorem}
Multi-partite $W$ state is uniquely determined by its single particle reduced density matrices. In other words,
for two given $W$-type states $\ket{\varphi},\ket{\psi}\in H_1\otimes H_2\otimes...\otimes
H_n$, if their reduced density matrices enjoys the same spectra, then $\ket{\varphi}_,\ket{\psi}$ are LU equivalent.
\end{theorem}
\textit{Proof}: Without loss of generality, assume that $\ket{\varphi},\ket{\psi}$ are given as
\begin{eqnarray*}
\ket{\varphi}=\sqrt{u}\ket{0\cdots 0}+\sqrt{a_1}\ket{0\cdots0 1}+\cdots+\sqrt{a_n}\ket{10\cdots 0},\\
\ket{\psi}=\sqrt{v}\ket{0\cdots 0}+\sqrt{b_1}\ket{0\cdots0 1}+\cdots+\sqrt{b_n}\ket{10\cdots 0}.
\end{eqnarray*}
their reduced density matrices satisfy $\det{\rho_k}=\det{\sigma_k}$ for all $k$, then the following holds for all $1\leq k\leq n$
$$a_k\sum_{j\neq k}^n a_j=\det{\rho_k}=\det{\sigma_k}=b_k\sum_{j\neq k}^n b_j.$$
One can obtain that $a_k=b_k$ for any $1\leq k\leq n$ by proving the following lemma.

Thus $u=1-\sum a_k=1-\sum b_k=v$, which means that $\ket{\varphi}=\ket{\psi}$.
\hfill $\blacksquare$

The following interesting lemma completes the proof of Theorem 1.
\begin{lemma}
$\{a_k:1\leq k\leq n\}$ and $\{b_k:1\leq k\leq n\}$ are two sets of positive numbers with $n\geq 3$, if
\begin{equation}
a_k\sum_{j\neq k}^n a_j =b_k\sum_{j\neq k}^n b_j ,
\end{equation}
is true for any $1\leq k\leq n$, then
$a_k$ = $b_k$ holds for any $1\leq k\leq n$.
\end{lemma}
\textit{Proof}: Consider (1) as equations of $\{a_k:1\leq k\leq n\}$, it is sufficient to show that
\begin{equation}
a_k\sum_{j\neq k}^n a_j =x_k/4,
\end{equation}
has at most one positive root, where $x_k=4 b_k\sum_{j\neq k}^n b_j$.

Let $A=\sum_{j=1}^n(a_j)$, we have
$$a_k=\frac{{A \pm \sqrt{A^{2}- x_k}}}{2}, ~~\mathrm{and}~~x_k=4 b_k\sum_{j\neq k}^n b_j=4 a_k(A-a_k) \leq (a_k+A-a_k)^2=A^2.$$
There is at most one $k$ such that $a_k\geq A/2$, that is
$a_k=\frac{{A + \sqrt{A^{2}- x_k}}}{2}$.
Without loss of generality, suppose the largest element of $\{a_k:1\leq k\leq n\}$ is $a_1$,
then for any $k\geq 2$, $a_k=\frac{{A-\sqrt{A^{2}- x_k}}}{2}$ and $a_1=A-\sum_{j=2}^n a_j$.

We only need to show that there is at most one solution, which satisfies Eq.(3) or Eq.(4), for given $x_1,x_2\cdots, x_n>0$.
\begin{eqnarray}
\sum_{k=1}^n(\frac{{A -\sqrt{A^{2}- x_k}}}{2})=A,\\
\frac{{A +\sqrt{A^{2}- x_1}}}{2}+\sum_{k=2}^n(\frac{{A -\sqrt{A^{2}- x_k}}}{2})=A,
\end{eqnarray}
Eq.(3) and Eq.(4) equals to $f(A)=0$ and $g(A)=0$ respectively, where
\begin{eqnarray*}
&f(y)&=-2y+\sum_{k=1}^n(\frac{x_k}{y+\sqrt{y^{2}-x_k}}),\\
&g(y)&=\sum_{k=2}^n(\sqrt{y^{2}-x_k})-(\sqrt{y^{2}-x_1}+(n-2)y).
\end{eqnarray*}

Case 1: Suppose there is some $r>0$ such that $f(r)=0$. It is direct to verify that $f(y)$ is a strictly monotone decreasing function, which implies that $f(y)$ has at most one root.

Assume $g(s)=0$ holds for some $s>0$.  Then $x_1\geq x_k$ for any $k$, otherwise $g(s)<0$ by noticing $s>\sqrt{s^2- x_k}$. Let $z=\sqrt{s^2-x_1}$, and for $k>1$ $z_k=\sqrt{x_1-x_k}$, then $0\leq z_k\leq \sqrt{x_1}$.

First, we show that $r=\sqrt{x_1}$. To do so, we suppose $r>\sqrt{x_1}$.
Since $f(y)$ is monotone decreasing, we have
$$0=f(r)< f(\sqrt{x_1})=-2\sqrt{x_1}+\sum_{i=1}^n\frac{x_i}{\sqrt{x_1}+\sqrt{x_1-x_i}}\Rightarrow \sum_{k=2}^n\sqrt{x_1-x_k} < (n-2) \sqrt{x_1}.$$
That is just $$\sum_{k=2}^n z_k<(n-2)\sqrt{x_1}.$$
Define real function for any $l>0$
$$h_l(y_2,y_3,\cdots,y_n)=\sum_{k=2}^n \sqrt{l^2+y_k^2},$$
Invoking the concavity of function $h_z(y_2,y_3\cdots,y_n)$, $0\leq z_k\leq \sqrt{x_1}$ and $\sum_{k=2}^n z_k<(n-2)\sqrt{x_1}$, we have the following
$$\sum_{k=2}^n \sqrt{s^{2}-x_k}=\sum_{k=2}^n \sqrt{z^2+z_k^2}=h_z(z_2,z_3,\cdots,z_n)<g(0,\sqrt{x_1},\cdots,\sqrt{x_1})=\sqrt{s^{2}-x_1}+(n-2)s.$$
But $g(s)=0$. Contradiction!

Thus, in this case, we know that $r=\sqrt{x_1}$. One can also obtain that $g(\sqrt{x_1})=0$ from $f(\sqrt{x_1})=0$.

Now, suppose there are $s_1>s_0>0$ such that $g(s_1)=g(s_0)=0$, assume $t=s_1^2-s_0^2$, we have
$$(n-2)s_1+\sqrt {s_1^2 - x_1}= \sum_{k=2}^n \sqrt{s_1^{2}-x_k}=\sum_{k=2}^n \sqrt{t+s_0^{2}-x_k}=h_{\sqrt{t}}(\sqrt{s_0^{2}-x_2},\sqrt{s_0^{2}-x_3},\cdots, \sqrt{s_0^{2}-x_n}).$$
According to $g(s_0)=0$, we have $\sum_{k=2}^n \sqrt{s_0^{2}-x_k} =(n-2)s_0+\sqrt {s_0^2 - x_1}$. Note that $\sqrt{s_0^{2}-x_k}<s_0$, we invoke the concavity of $h_z(y_2,y_3\cdots,y_n)$ again and obtain that
$$h_{\sqrt{t}}(\sqrt{s_0^{2}-x_2},\sqrt{s_0^{2}-x_3},\cdots, \sqrt{s_0^{2}-x_n})<h_{\sqrt{t}}(0,s_0,\cdots, s_0)=(n-2)s_1+\sqrt {s_1^2 - x_1},$$
which is a contradiction from $g(s_1)=0$.

Therefore, $g(y)$ can have at most one root.

Thus, $f(y)$ has at most one root. If $f(r)=0$ and $g(s)=0$, $r=s=\sqrt{x_1}$.

Case 2: Suppose $f(y)$ has no root. As we discussed above, $g(y)$ has at most one root.

Thus, there is at most one solution, which satisfies Eq.(3) or Eq.(4).

Obviously, we can choose $A=\sum_{k=1}^n b_k$, then $a_k=b_k$ is the only possible case. \hfill $\blacksquare$

In this note, we show that the entanglement of $W-$type states are uniquely determined by its reduced density matrices.

I thank Dr R. Duan, C. Guo, L. Chen for very helpful discussion. I was indebted to M. Ying for his constant support. This work was partly supported by the National Natural Science Foundation of China
(Grant Nos. 61179030 and 60621062), the Australian Research Council (Grant
Nos. DP110103473 and DP120103776).

This work was accomplished in 2010, and recently, we have learned about the independent work by Sawicki, Walter, Kus \cite{SWK12}, in which the three-qubit case is studied and they left the conjecture of the n-qubit $W-$type state. Our result provides a positive answer to that conjecture.

\end{document}